\documentclass{aa}
\usepackage{graphicx}
\usepackage{txfonts}
\usepackage{natbib}
\begin{document}
   \title{XMM-Newton X-ray and optical observations of the globular clusters M~55 and NGC~3201}

   \subtitle{}

   \author{N.A. Webb
          \inst{1}
          \and
          P.J. Wheatley
          \inst{2}
          \and
          D. Barret
          \inst{1}
          }

   \institute{Centre d'Etude Spatiale des Rayonnements, 9 avenue du
             Colonel  Roche, 31028 Toulouse Cedex 04, France \and
             Department of Physics and Astronomy, University of
             Leicester,  Leicester LE1 7RH, England }

   \offprints{N.A. Webb \email{Natalie.Webb@cesr.fr}}

   \date{Received 2005; accepted 2005}

   \abstract{We have observed two low concentration Galactic globular
   clusters with the X-ray observatory {\it XMM-Newton}.  We detect 47
   faint X-ray sources in the direction of \object{M 55} and 62 in the
   field of view of \object{NGC 3201}.  Using the statistical
   Log~N-Log~S relationship of extragalactic sources derived from {\it
   XMM-Newton} Lockman Hole observations, to estimate the background
   source population, we estimate that very few of the sources
   (1.5$\pm$1.0) in the field of view of \object{M 55} actually belong
   to the cluster.  These sources are located in the centre of the
   cluster as we expect if the cluster has undergone mass
   segregation. \object{NGC 3201} has approximately 15 related
   sources, which are centrally located but are not constrained to lie
   within the half mass radius. The sources belonging to this cluster
   can lie up to 5 core radii from the centre of the cluster which
   could imply that this cluster has been perturbed.  Using X-ray (and
   optical, in the case of \object{M 55}) colours, spectral and timing
   analysis (where possible) and comparing these observations to
   previous X-ray observations, we find evidence for sources in each
   cluster that could be cataclysmic variables, active binaries,
   millisecond pulsars and possible evidence for a quiescent low mass
   X-ray binary with a neutron star primary, even though we do not
   expect any such objects in either of the clusters, due to their low
   central concentrations.  The majority of the other sources are
   background sources, such as AGN. \footnote{Tables 2 and 3 are
   available in electronic form at the CDS via anonymous ftp to
   cdsarc.u-strasbg.fr (130.79.128.5) or via
   http://cdsweb.u-strasbg.fr/cgi-bin/qcat?J/A+A/}

    }

\authorrunning{Webb et al.}  \titlerunning{X-ray/optical sources in M
55 and NGC 3201}

   \maketitle
%

\section{Introduction}

Following the discovery, over twenty years ago, of low luminosity
(L$_x\ _\sim ^<$ 10$^{34.5}$ erg s$^{-1}$) X-ray sources in globular
clusters \citep{hert83}, the known population of such sources has
steadily grown \citep[e.g.][]{john94,cool95,john96,geff97,verb01}.
However, with the launch of the X-ray observatories {\it XMM-Newton}
and {\it Chandra}, whose sensitivity and angular resolution
(respectively) far surpasses those of previous X-ray observatories, the
number of new sources detected (and identified) has truly flourished
\citep[e.g.][]{webb04c,gend03a,pool03,hein03,gend03b,rutl02,webb01,grin01}.
The low luminosity X-ray sources that belong to their respective
clusters are mostly a variety of binary systems (X-ray binaries,
cataclysmic variables and active binaries) as well as descendants of
these objects (such as millisecond pulsars).  These X-ray sources are
thus indispensable objects with which to complete the study of stellar
and globular cluster evolution, stellar dynamics, binary formation and
evolution, as well as the study of the binaries themselves, see e.g.
\cite{verb04,hut92} for reviews of these topics.

We have thus observed two further globular clusters (\object{M 55}
(\object{NGC 6809}) and \object{NGC 3201}) using {\it XMM-Newton}, to
detect their populations of low luminosity X-ray sources and determine
the nature of these sources, as well as probe the evolution of the
clusters.

The two globular clusters presented in this paper are particularly
interesting to study as they have very low central concentrations of
stars.  It is thought that through dynamical evolution, cluster cores
become increasingly concentrated \citep{heno61}.  Stellar
encounters in globular clusters result in an equipartition of energy.
Through virialization, the more massive stars of the cluster, such as
degenerate objects and binary systems, are concentrated towards the
centre of the cluster.  Low concentration clusters are therefore least
affected by dynamical processes and thus mass segregation is thought
to be less important for these clusters than the higher concentration
clusters.  These clusters are also particularly susceptible to tidal
shocking \citep{cher86} and to a loss of stars from the cluster.

Both clusters have been studied at long and short wavelengths.
\object{M 55} has previously been observed in the X-ray domain by the
X-ray telescopes {\it Ariel V} \citep{pye83}, {\it EINSTEIN}, {\it
HEAO 1} \citep{hert85} and {\it Rosat}
\citep{john96,verb01}. \cite{pye83} found a transient source with {\it
Ariel V}, that lies towards the edge of the cluster (outside the field of view of our
observations).  \cite{john96} found 18 sources with the {\it Rosat}
PSPC.  A number of studies in the optical domain have analysed the
nature of the cluster through the study of its luminosity function and
the distribution of stars, blue stragglers, binary systems and
variable stars e.g. \cite{piot99,zagg97,mand97,pych01,olec99}.

\object{NGC 3201} has also been studied with the X-ray telescopes
{\it HEAO 1} \citep{hert85} and {\it Rosat} \citep{john96,verb01}.
Nine sources were found by \cite{john96} and an additional one by
\cite{verb01}.  This cluster has also been extensively studied in the
optical domain e.g. \cite{layd03,mazu03,cove03,pier02,vonb02}.

\section{Observations and data reduction}

\subsection{X-ray data}

We obtained {\it XMM-Newton} observations of the two globular clusters
\object{M 55} and \object{NGC 3201}, during the `Routine Observing
Phase'.  The observations are given in Table~\ref{tab:obs}.
Approximately 3 and 4 ks of the {\it pn} data of the two clusters
respectively were affected by high background activity (a soft proton
flare).  All of the data were reduced with version 6.0 of the {\it
XMM-Newton} SAS (Science Analysis Software).

The MOS data were reduced using the `emchain' with `embadpixfind' to
detect the bad pixels.  The event lists were filtered, so that 0-12 of
the predefined patterns (single, double, triple, and quadruple pixel
events) were retained and the high background periods were identified
by defining a count rate threshold above the low background rate and
the periods of higher background counts were then flagged in the event
list.  We also filtered in energy. We used the energy range 0.2-10.0
keV, as recommended in the document `EPIC Status of Calibration and
Data Analysis' \citep{kirs02}.

The {\it pn} data were reduced using the `epchain' of the SAS.  Again
the event lists were filtered, so that 0-4 of the predefined patterns
(single and double events) were retained, as these have the best
energy calibration.  We again filtered in energy, where we used the
energy range 0.5-10.0 keV.

\begin{table}[!h]
  \caption[]{Summary of the observations presented in this paper}
     \label{tab:obs}
       \begin{tabular}{lcccc}
         \hline \noalign{\smallskip} Cluster/ & Instru- & Filter &
         Mode & Useful  \\ Date     &   ment               &      &  &
         exposure (ks)\\ \hline M 55/ & MOS & medium & Full frame$^*$
         & 26 \\ 17-18/10/01 & pn  & medium & Full frame & 24  \\ & OM
         & UVW2, B  & Imaging   &2$\times$1, 1\\   \hline NGC 3201/ &
         MOS & medium & Full frame & 40 \\ 25/05/03 & pn  & medium &
         Full frame & 42 \\ & OM  & UVW1  & Imaging    &2$\times$4.82,
         0.8\\   \hline
  \end{tabular}
\vspace*{-0.3cm}     $^*$ \cite{turn01}
 \end{table}

\begin{figure}[!t]
   \centering \includegraphics[angle=0,width=8.5cm]{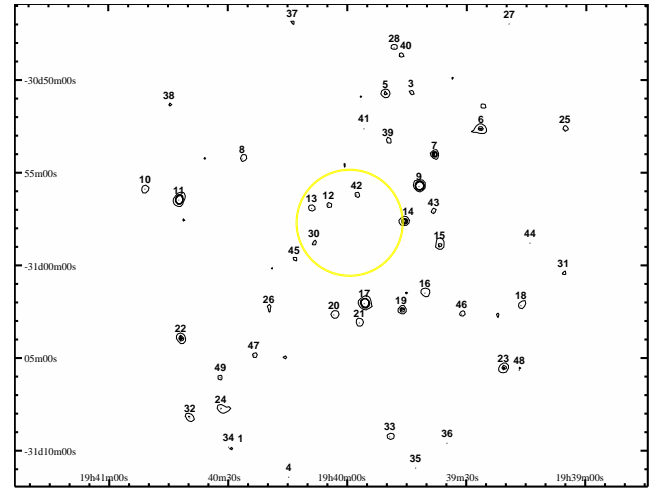}
\caption{X-ray contour plot of the central region of M~55. The circle
   shows both the core and the half-mass radii which are almost
   coincident.  The abscissa indicates the right ascension in $^h$
   \hspace*{1mm} $^m$
   \hspace*{1mm} $^s$ and the ordinate indicates the declination in
   $^{\circ}$ \hspace*{1mm} \arcmin \hspace*{1mm} \arcsec.  The
   contours represent 4, 6 and 8 $\sigma$ confidence levels and the
   source numbers are those found in Table~\ref{tab:xsourcesm55}. }
\label{fig:m55contours}
\end{figure}

\begin{figure}[!t]
   \centering \includegraphics[angle=0,width=8.5cm]{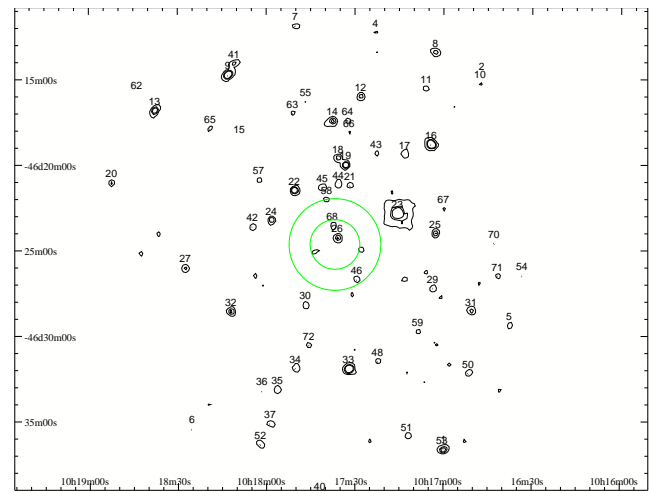}
\caption{X-ray contour plot of the central region of NGC~3201
        (similar to Fig.~\ref{fig:m55contours}). The source numbers
        are those found in Table~\ref{tab:xsourcesngc3201}. Source 23
        looks slightly extended in this combined image, but it is a
        point source.}
\label{fig:ngc3201contours}
\end{figure}

The source detection was done in the same way as
\cite{gend03a,webb04c}, but in summary we used the SAS wavelet
detection algorithm on the 0.5-5.0 keV image, as the signal-to-noise
ratio was best in this range.  We kept only those sources detected
with two or more cameras, with the exception of sources found with the
{\it pn}  camera, which is more sensitive than the two MOS cameras.
We detected 47 sources in the field of view (FOV)
(radius$\sim$15\arcmin) of \object{M 55} and 62 sources in the FOV of
\object{NGC 3201}, with a maximum likelihood greater than 4.5$\sigma$.
These sources are given in Tables~\ref{tab:xsourcesm55} and
\ref{tab:xsourcesngc3201}, along with their position and count rate.
We also give the identification number for the sources detected by
{\it Rosat}.  The positional error is the 90\% confidence mean
statistical error on the position of the {\it XMM-Newton} sources.
The majority of these sources can be seen in the contour plots of the
central regions of these globular clusters, see
Figs.~\ref{fig:m55contours} and \ref{fig:ngc3201contours}.

\begin{table}[!h]
  \caption[]{X-ray sources in the direction of M~55, as determined
  from the EPIC observations, 0.5-10.0 keV band.  The identification
  number, any former identifications and position are given, along
  with the error on this position and the count rate.}
     \label{tab:xsourcesm55}
       \begin{tabular}{lccccc}
         \hline \noalign{\smallskip} & For. & RA (2000) & Dec (2000) &
         Error & Count s$^{-1}$ \\ ID & ID & $^h$ \hspace*{3mm} $^m$
         \hspace*{3mm} $^s$ & $^{\circ}$ \hspace*{3mm} ' \hspace*{3mm}
         '' & '' & $\times10^{-3}$ \\ \hline 1& & 19\ 40\ 29.04 & -31\
         9\ 54.36 & 7.96 & 3.96$\pm$0.95 \\ 3& & 19\ 39\ 44.00 & -30\
         50\ 42.98 & 6.74 & 3.85$\pm$0.76 \\ 4& & 19\ 40\ 14.76 & -31\
         11\ 27.67 & 6.78 & 8.09$\pm$1.30 \\ 5& & 19\ 39\ 50.40 & -30\
         50\ 47.18 & 4.98 & 10.87$\pm$1.21 \\ 6& & 19\ 39\ 26.38 &
         -30\ 52\ 40.32 & 4.88 & 16.69$\pm$1.52 \\ 7& & 19\ 39\ 38.14
         & -30\ 54\ 4.60 &  4.52 & 10.92$\pm$0.91 \\ 8& & 19\ 40\
         26.43 & -30\ 54\ 16.95 & 6.23 &  4.84$\pm$0.73 \\ 9& 6 & 19\
         39\ 41.92&-30\ 55\ 45.88 & 4.20 & 26.30$\pm$1.34 \\ 10& 7 &
         19\ 40\ 50.85&-30\ 55\ 54.31 & 5.22 & 13.72$\pm$1.69 \\ 11& 8
         & 19\ 40\ 42.35&-30\ 56\ 30.12 & 4.22 & 50.59$\pm$2.55 \\
         12$^c$& & 19\ 40\ 4.75  & -30\ 56\ 46.84 & 9.60 &
         1.27$\pm$0.31 \\ 13$^c$& & 19\ 40\ 8.96  & -30\ 56\ 57.26 &
         5.91 & 3.34$\pm$0.59 \\ 14$^h$& & 19\ 39\ 45.77 & -30\ 57\
         40.14 & 4.46 & 10.44$\pm$0.98 \\ 15& & 19\ 39\ 36.78 & -30\
         58\ 56.86 & 4.80 & 8.24$\pm$0.80 \\ 16& & 19\ 39\ 40.44 &
         -31\  1\ 31.48 & 4.96 & 7.43$\pm$0.85 \\ 17& 13 & 19\ 39\
         55.69&-31\ 2\ 5.01  & 4.17 & 31.60$\pm$1.49 \\ 18& & 19\ 39\
         16.28 & -31\  2\ 11.08 & 5.08 & 7.28$\pm$0.97 \\ 19& & 19\
         39\ 46.46 & -31\  2\ 26.91 & 4.60 & 8.40$\pm$0.84 \\ 20& &
         19\ 40\ 3.36  & -31\ 2\ 41.65  & 5.13 & 6.29$\pm$0.82\\ 21& &
         19\ 39\ 56.94 & -31\  3\ 7.61  & 4.87 & 5.83$\pm$0.72 \\ 22&
         15 & 19\ 40\ 41.98 &-31\ 3\ 58.28& 4.54 & 21.08$\pm$1.84 \\
         23& & 19\ 39\ 20.68 & -31\  5\ 33.77 & 4.78 & 15.68$\pm$1.57
         \\ 24& & 19\ 40\ 31.87 & -31\ 7\ 45.96  & 5.01 &
         17.20$\pm$1.90 \\ 25& 5 & 19\ 39\ 5.34& -30\ 52\ 41.18 & 5.97
         & 7.94$\pm$1.32 \\ 26& & 19\ 40\ 19.76 & -31\ 2\ 25.13
         &11.44 & 1.43$\pm$0.37 \\ 27& & 19\ 39\ 19.35 & -30\ 46\
         59.67 & 7.58 & 3.76$\pm$0.89 \\ 28& & 19\ 39\ 48.21 & -30\
         48\ 17.35 & 5.70 & 6.29$\pm$1.14\\ 30$^c$& 9 & 19\ 40\ 8.40
         &-30\ 58\ 48.82 & 6.98 & 1.40$\pm$0.30 \\ 31& 12 & 19\ 39\
         5.63 &-31\ 0\ 25.46 & 6.94 & 2.34$\pm$0.53 \\ 32& & 19\ 40\
         39.67 & -31\ 8\  15.00 & 4.82 & 20.02$\pm$2.26 \\ 33& & 19\
         39\ 49.26 & -31\  9\ 16.30 & 5.70 & 8.56$\pm$1.25 \\ 34& &
         19\ 40\ 29.88 & -31\ 9\ 50.29  & 7.21 & 4.00$\pm$0.87 \\ 35&
         & 19\ 39\ 42.75 & -31\ 10\ 57.82 & 7.45 & 2.17$\pm$0.59 \\
         36& & 19\ 39\ 34.83 & -31\ 9\ 37.49 & 10.97 & 3.24$\pm$0.89
         \\ 37& & 19\ 40\ 13.80 & -30\ 46\ 56.34 & 7.83 &
         4.37$\pm$1.02 \\ 38& & 19\ 40\ 44.85 & -30\ 51\ 21.35 & 9.39
         & 5.07$\pm$1.41 \\ 39& & 19\ 39\ 49.82 & -30\ 53\ 21.63 &
         8.04 & 1.98$\pm$0.52 \\ 40& & 19\ 39\ 46.74 & -30\ 48\ 42.22
         & 7.96 & 3.94$\pm$0.91 \\ 41& & 19\ 39\ 55.79 & -30\ 52\
         39.14 & 7.26 & 1.74$\pm$0.43 \\ 42$^c$& & 19\ 39\ 57.64 &
         -30\ 56\ 13.82 & 8.52 & 2.11$\pm$0.47 \\ 43& & 19\ 39\ 38.26
         & -30\ 57\ 8.22 & 7.52  & 2.15$\pm$0.49 \\ 44& & 19\ 39\
         14.01 & -30\ 58\ 49.21 & 7.98 & 2.00$\pm$0.60 \\ 45& & 19\
         40\ 13.380& -30\ 59\ 43.71 & 7.33 & 1.60$\pm$0.43 \\ 46& &
         19\ 39\ 31.25 & -31\ 2\ 39.44 &  6.25 & 2.28$\pm$0.52 \\ 47&
         & 19\ 40\ 23.55 &  -31\ 4\ 52.85 & 7.98 & 3.07$\pm$0.81 \\
         48& & 19\ 39\ 16.66 & -31\ 5\ 41.05 & 6.88 &  3.34$\pm$0.79\\
         49& 16 & 19\ 40\ 31.96&-31\ 6\ 4.79 & 8.00 &  4.23$\pm$1.02
         \\ \hline
  \end{tabular}
    \begin{list}{}{}
\vspace*{-0.3cm}
      \item $^c$ sources within the core, $^h$ sources within the half
      mass radius
      \item \hspace*{1.0cm}numbers, {\it Rosat} PSPC, \cite{john96}
    \end{list}
\vspace*{-0.3cm}
 \end{table}

\subsection{Optical data}

The {\it XMM-Newton} optical monitor \citep{maso01} was also employed
during the X-ray observations and the data obtained is given in
Table~\ref{tab:obs}.  The data were reduced using the standard SAS
{\it omichain} script.

We find 345 stars with a minimum significance of 3.0 in the FOV of
\object{M 55}, detected in both the UVW2- and the B-band.  The
magnitudes range from approximately 14.0 to 19.5 in the UVW2-filter
and 12.0 to 20.5 in the B-filter. The photometric errors are poor for
the UVW2-filter ranging from 0.001 to 1.5 magnitudes with a mean value
of 0.22, but better for the B-filter, ranging from 0.001-0.6
magnitudes with a mean value of 0.03.

We find 7569 stars with a minimum significance of 3.0 in the FOV of
\object{NGC 3201}.  The magnitudes range from approximately 12  to
22 with photometric errors ranging from 0.001 to 1.5 magnitudes with a
mean value of 0.17.  However, as the observations have been made using
a single filter (UVW1), these data are of limited use in determining
the optical counterparts to the X-ray sources.

\section{The X-ray sources}
\label{sec:xsources}

\subsection{M 55}
\label{sec:m55xsources}

We have detected four sources within the core radius
\citep[2.83\arcmin, centred at  RA=19$^h$ 39$^m$ 59.40$^s$, dec=-30$^{\rm \circ}$ 57\arcmin\ 43.5\arcsec][]{djor93}
of \object{M 55}, sources 12, 13, 30, and 42 (indicated by a $^c$ in
Table~\ref{tab:xsourcesm55}), and a further source (14) that lies just
on the limit of the half-mass radius \citep[2.89\arcmin][]{harr99}
(indicated by a $^h$ in Table~\ref{tab:xsourcesm55}).  We computed the
detectable flux limit within the half mass radius, assuming a
0.6 keV blackbody spectral model in the same way as
\citet{john96a}. The limiting luminosity at the centre of the field of
view for the {\it pn} camera is L$_x$ = $7 \times 10^{30}$ erg
s$^{-1}$ (0.5-10.0 keV) and L$_x$ = $2 \times 10^{31}$ erg s$^{-1}$
(0.5-10.0 keV) for the MOS.

Many of our detected sources are background sources. We have used the
statistical Log~N-Log~S relationship of extragalactic sources derived
from the Lockman Hole observations \citep{hasi01} to estimate the
background source population, assuming that there is no cosmic
variance. We converted the source count rates to fluxes using a power
law model with a spectral index of -2, as \cite{hasi01}. We employed
the method of \cite{cool02} and \cite{gend03a} which takes into
account vignetting, as the faintest detectable flux increases
with radius. We calculated the average limiting flux in three annuli
on the {\it pn}  camera, in the same way as \citet{john96a}.
These annuli were bound by the core/half-mass radii, three and five
core radii.  We expect 3.5$\pm$1, 21.2$\pm$2 and 22.3$\pm$5 background
sources in these annuli respectively.  The uncertainty is the 10\%
error estimated on the flux, as \cite{hasi01}. We find 5, 17 and 25
sources in these annuli respectively.  Although the sample is small
and therefore the results are not very robust, this indicates that at
least one of the core/half-mass radii sources is likely to belong to
the cluster.  Using the 5 sources found in the half mass radius and
the fact that approximately 1.5 may belong to the cluster, we
calculate using Poisson statistics, that the probability that none of
the 5 sources are cluster members is 27\%. This supports the idea that
not all the core sources belong to the cluster.

It is also possible to calculate the probability that the sources
within the core are not simply spurious identifications with the
cluster, in the same manner as \cite{verb01}.  The  probability (p)
that  one serendipitous  source in  the {\it pn}  observation  is at a
distance  $r < R$ to  the cluster centre, where $R$ is the  core
radius, is $p=(R/r_d)^2$, where $r_d$ is the radius of the field of
view. The probability of a randomly located source falling within the
core radius is 3.6 percent.  For the 47 sources, the probability of
finding no sources within the core radius by chance is (1-p)$\times$47 = 18\%.
This again supports the assertion that not all the core sources belong
to the cluster.

\begin{table}
  \caption[]{As Table~\ref{tab:xsourcesm55} but in the direction of
  NGC~3201.}
     \label{tab:xsourcesngc3201}
       \begin{tabular}{lccccc}
         \hline \noalign{\smallskip} & For. & RA (2000) & Dec (2000) &
         Error & Count s$^{-1}$ \\ ID & ID & $^h$ \hspace*{3mm} $^m$
         \hspace*{3mm} $^s$ & $^{\circ}$ \hspace*{3mm} ' \hspace*{3mm}
         '' & '' & $\times10^{-3}$ \\ \hline 1& & 10\ 16\ 59.90 & -46\
         36\ 38.90 & 4.94 & 14.12$\pm$1.27\\ 2& & 10\ 16\ 47.23 & -46\
         14\ 49.24 & 7.67 & 1.48$\pm$0.37\\ 4& & 10\ 17\ 23.13 & -46\
         12\ 18.22 & 5.49 & 6.12$\pm$1.19\\ 5& & 10\ 16\ 37.66 & -46\
         29\ 24.73 & 5.58 & 6.50$\pm$0.99\\ 6& & 10\ 18\ 25.40 & -46\
         35\ 28.63 & 6.17 & 4.39$\pm$1.04\\ 7& & 10\ 17\ 50.21 & -46\
         11\ 55.97 & 6.36 & 8.72$\pm$1.27\\ 8& & 10\ 17\ 2.89  & -46\
         13\ 29.03 & 4.84 & 16.62$\pm$1.55\\ 9& 3 & 10\ 18\ 13.00 &
         -46\ 14\ 47.19 & 4.24 & 49.28$\pm$2.49\\ 10& & 10\ 16\ 47.76
         & -46\ 15\ 16.54 & 7.34 & 1.85$\pm$0.49\\ 11& & 10\ 17\ 5.97
         & -46\ 15\ 36.15 & 5.02 & 7.07$\pm$1.17\\ 12& & 10\ 17\ 28.11
         & -46\ 16\ 3.52  & 4.97 & 10.00$\pm$1.04\\ 13& 4 & 10\ 18\
         37.76 & -46\ 16\ 52.53 & 4.33 & 39.19$\pm$2.62\\ 14& & 10\
         17\ 37.70 & -46\ 17\ 29.72 & 4.51 & 13.27$\pm$1.01\\ 15& &
         10\ 18\ 9.13  & -46\ 18\ 32.61 & 8.37 & 1.83$\pm$0.46\\ 16& &
         10\ 17\ 4.28  & -46\ 18\ 50.88 & 4.17 & 35.43$\pm$1.61\\ 17&
         & 10\ 17\ 13.26 & -46\ 19\ 25.62 & 5.02 & 6.13$\pm$0.70\\ 18&
         & 10\ 17\ 35.76 & -46\ 19\ 39.84 & 4.70 & 6.94$\pm$0.72\\ 19&
         & 10\ 17\ 33.26 & -46\ 20\ 3.51  & 4.28 & 13.74$\pm$0.88\\
         20& & 10\ 18\ 52.44 & -46\ 21\ 5.50  & 4.86 &
         22.21$\pm$2.06\\ 21& & 10\ 17\ 31.72 & -46\ 21\ 15.74 & 5.95
         & 2.53$\pm$0.41\\ 22& & 10\ 17\ 50.54 & -46\ 21\ 32.90 & 4.25
         & 15.75$\pm$0.96\\ 23& 6 & 10\ 17\ 15.63 & -46\ 22\ 54.06 &
         4.04 & 87.13$\pm$1.55\\ 24& & 10\ 17\ 58.33 & -46\ 23\ 17.00
         & 4.55 & 8.07$\pm$0.71\\ 25& & 10\ 17\ 2.85  & -46\ 24\ 3.93
         & 4.65 & 8.07$\pm$0.71\\ 26$^c$& C & 10\ 17\ 36.00 & -46\ 24\
         19.50 & 4.57 & 7.09$\pm$0.58\\ 27& & 10\ 18\ 27.54 & -46\ 26\
         3.48  & 4.97 & 10.18$\pm$1.08\\ 29& & 10\ 17\ 3.64  & -46\
         27\ 15.52 & 5.37 & 3.44$\pm$0.53\\ 30& & 10\ 17\ 46.85 & -46\
         28\ 15.00 & 4.99 & 4.13$\pm$0.53\\ 31& & 10\ 16\ 50.58 & -46\
         28\ 35.89 & 4.81 & 9.21$\pm$0.88\\ 32& & 10\ 18\ 12.08 & -46\
         28\ 36.49 & 4.60 & 12.63$\pm$1.0\\ 33& & 10\ 17\ 32.20 & -46\
         31\ 57.59 & 4.27 & 24.90$\pm$1.35\\ 34& & 10\ 17\ 50.25 &
         -46\ 31\ 57.61 & 5.19 & 5.96$\pm$0.80\\ 35& & 10\ 17\ 56.34 &
         -46\ 33\ 11.57 & 5.01 & 7.34$\pm$0.89\\ 36& & 10\ 18\ 1.57  &
         -46\ 33\ 15.41 & 6.68 & 3.81$\pm$0.75\\ 37& & 10\ 17\ 58.81 &
         -46\ 35\ 11.47 & 5.11 & 8.80$\pm$1.12\\ 40& & 10\ 17\ 41.92 &
         -46\ 39\ 22.59 & 10.11& 2.45$\pm$0.62\\ 41& & 10\ 18\ 10.89 &
         -46\ 14\ 6.96  & 5.00 & 12.81$\pm$1.39\\ 42& & 10\ 18\ 4.76 &
         -46\ 23\ 40.79 & 5.57 & 4.40$\pm$0.61\\ 43& & 10\ 17\ 22.86 &
         -46\ 19\ 23.53 & 6.36 & 2.34$\pm$0.45\\ 44& & 10\ 17\ 35.68 &
         -46\ 21\ 11.29 & 5.27 & 4.08$\pm$0.55\\ 45& & 10\ 17\ 41.08 &
         -46\ 21\ 22.55 & 5.18 & 3.96$\pm$0.56\\ 46$^h$& & 10\ 17\
         29.42 & -46\ 26\ 44.96 & 5.49 & 2.98$\pm$0.45\\ 48& & 10\ 17\
         22.36 & -46\ 31\ 34.01 & 5.49 & 3.29$\pm$0.51\\ 50& & 10\ 16\
         51.55 & -46\ 32\ 13.14 & 5.75 & 5.79$\pm$0.85\\ 51& & 10\ 17\
         12.40 & -46\ 35\ 54.46 & 5.40 & 6.54$\pm$0.98\\ 52& & 10\ 18\
         2.11  & -46\ 36\ 22.69 & 5.39 & 10.80$\pm$1.41\\ 53& & 10\
         17\ 0.35  & -46\ 36\ 42.83 & 4.96 & 10.63$\pm$1.09\\ 54& &
         10\ 16\ 33.37 & -46\ 26\ 30.55 & 8.78 & 3.26$\pm$0.76\\ 55& &
         10\ 17\ 46.74 & -46\ 16\ 20.56 & 8.53 & 2.44$\pm$0.62\\ 57& &
         10\ 18\ 2.61  & -46\ 20\ 54.86 & 8.38 & 1.99$\pm$0.46\\
         58$^h$& & 10\ 17\ 39.60 & -46\ 22\ 3.71  & 8.03 &
         1.46$\pm$0.35\\ 59& & 10\ 17\ 8.71  & -46\ 29\ 47.34 & 8.67 &
         1.72$\pm$0.45\\ 62& & 10\ 18\ 43.83 & -46\ 15\ 57.42 & 5.21 &
         13.16$\pm$1.74\\ 63& & 10\ 17\ 51.18 & -46\ 17\ 3.33  & 6.95
         & 2.59$\pm$0.59\\ 64& & 10\ 17\ 32.48 & -46\ 17\ 29.54 & 8.04
         & 1.74$\pm$0.51\\ 65& & 10\ 18\ 19.01 & -46\ 17\ 55.49 & 6.71
         & 3.88$\pm$0.82\\ 66& & 10\ 17\ 32.07 & -46\ 18\ 10.60 & 7.40
         & 1.45$\pm$0.40\\ 67& & 10\ 17\ 0.38  & -46\ 22\ 36.30 & 7.31
         & 2.09$\pm$0.50\\ 68$^c$& & 10\ 17\ 37.72 & -46\ 23\ 36.30 &
         9.90 & 1.43$\pm$0.37\\ 70& & 10\ 16\ 42.89 & -46\ 24\ 37.22 &
         7.36 & 2.09$\pm$0.6\\ 71& & 10\ 16\ 41.75 & -46\ 26\ 32.46 &
         6.89 & 2.85$\pm$0.63\\ 72& & 10\ 17\ 45.66 & -46\ 30\ 35.88 &
         7.24 & 1.76$\pm$0.45\\ \hline
  \end{tabular}
 \end{table}

\subsection{NGC 3201}
\label{sec:ngc3201xsources}

\object{NGC 3201}, centred at RA=10$^h$ 17$^m$ 36.8$^s$,
dec=-46$^{\circ}$ 24\arcmin\ 40\arcsec\ \citep{harr99}, has fewer
centrally located X-ray sources. Within the core radius
\citep[1.45\arcmin][]{djor93}, we find only two sources, sources 26
and 68 (indicated by a $^c$ in Table~\ref{tab:xsourcesngc3201}).  We
find another two in the half-mass radius
\citep[2.68\arcmin][]{harr99}, sources 46 and 58 (indicated by a $^h$
in Table~\ref{tab:xsourcesm55}).  The limiting luminosity for the {\it
pn}  camera is L$_x$ = $8 \times 10^{30}$ erg s$^{-1}$ (0.5-10.0 keV)
and L$_x$ = $2 \times 10^{31}$ erg s$^{-1}$ (0.5-10.0 keV) for the MOS
camera.

To determine the number of expected background sources, we defined six
annuli bound by the core radius, the half-mass radius, 3, 5, 7.5 and
10 core radii.  We expect 0.7$\pm$0.1, 1.8$\pm$0.4 and 3.3$\pm$0.7,
7.3$\pm$2.0, 13.8$\pm$2.0 and 18.4$\pm$2.0 background sources in these
annuli. We find 2, 2, 7, 17, 15 and 19 sources in these annuli
respectively.  This indicates an over-density of sources in almost
every annulus, except the outer two, which is different to many other
globular clusters and is discussed in Section~\ref{sec:discuss}.
Using the 4 sources found in the half mass radius and the fact that
approximately 1.5 may belong to the cluster, we calculate using
Poisson statistics, that the probability that none of the 4 sources
are cluster members is 24\%. This supports the idea that not all the
core sources belong to the cluster. The probability of a randomly
located source not falling in the core radius is 99.0\%.  For the 62
sources, the probability of finding no sources within the core radius
by chance is 55\%.  It is therefore likely that at least one of the
core sources does belong to the cluster.

However, as stated above, we find an excess of sources in many of the
defined annuli.  Considering the four innermost annuli, we find 28
sources where we expect approximately 13.  Poisson statistics indicate
that the probability that none of these sources belong to the cluster
is 0.02\%.  This implies that at least one of the sources does belong
to the cluster.

\subsection{Spectral analysis}
\label{sec:xspect}

Seven X-ray sources in \object{M 55} and nine sources in \object{NGC
3201} have enough counts that we could extract and fit the spectra.
We extracted the spectra using circles of radii $\sim$1\arcmin\
centred on the source, with the exception of source 19 in \object{NGC
3201} where we used an extraction radius of $\sim$30\arcsec\ due its
close proximity to other sources.  We used a similar neighbouring
region, free from X-ray sources to extract a background file.  We
rebinned the MOS data into 15 eV bins and the {\it pn}  data into 5eV
bins.  We used the SAS tasks `rmfgen' and `arfgen' to generate a
`redistribution matrix file' and an `ancillary response file', for
each source.  We binned up the data to contain at least 20 net
counts/bin.  We then used Xspec (Version 11.2.0) to fit the spectra.
We initially tried simple models, such as a power law, a blackbody, a
bremsstrahlung and a Raymond Smith fit, allowing the absorption 
and the model parameters to vary freely, so that the best
fit values obtained do not depend on the initial parameters chosen for
the model.  If the fit could not be constrained with the freely
varying absorption value, we fixed the absorption value to that
of the cluster.  We found that simple models provided a good fit to
the data for the majority of the sources.  However, we found that for
certain sources, we required a more complicated model to fit the data.
The results of the spectral fitting can be found in
Tables~\ref{tab:xspectram55} and \ref{tab:xspectrangc3201}, where the
best fits are given, along with the goodness of fit
($\chi^{\scriptscriptstyle 2}_{\scriptscriptstyle \nu}$) and the flux
in the 0.2-10.0 keV range.

\begin{table*}
\begin{minipage}{18cm}
\caption{In the left hand side of the table the best fitting models to
spectra from the MOS and {\it pn}  data (with the exception of source
14, which uses only the MOS data as the source fell in a chip gap in
the {\it pn}  data and source 6, which uses only the {\it pn}  data as
the source is too faint in the MOS), for the sources in \object{M 55}.
Where no error value is given for the N$_H$ ($\times 10^{21}{\rm
cm}^{-2}$), the N$_H$ was frozen to that of the cluster
\citep[0.39$\times 10^{21}{\rm cm}^{-2}$][]{harr99,pred95}.  The flux
given is unabsorbed flux ($\times 10^{-13} {\rm ergs\ cm}^{-2} {\rm
s}^{-1}$, 0.2-10.0 keV), with errors of the order $\pm$10\%.  The
right hand side of the table contains the results from testing for
variability.  This includes the number of data bins, the length (in
seconds) of each data bin and the probabilities from a
Kolmogorov-Smirnov probability of constancy test and the $\chi^2$
probability of constancy test.  Source numbers correspond to those
given in Table~\ref{tab:xsourcesm55} and $^m$ indicates possible
cluster members, see Sect.~\ref{sec:discuss}.}
\label{tab:xspectram55}
\begin{center}
\begin{tabular}{cccccccccc|ccc}
\hline Src & N$_H$  & Model & kT (keV) & Photon & Abundance & z &
$\chi^{\scriptscriptstyle 2}_{\scriptscriptstyle \nu}$ & dof & Flux &
bins & time & KS/ \\ 
& $\times 10^{21}{\rm cm}^{-2}$ & & &Index  & & &
& &  & & (s) & $\chi^2$ \\ 
\hline 
\hline 
14 & 45.82$\pm$37.02 & PL & -
& 3.67$\pm$2.28 & - & & 0.88 & 13 & 22.0 & 25404 & 1 &
4$\times10^{-8}$/\\ 
& 27.17$\pm$22.78 & BB & 0.64$\pm$0.27 & - & - & & 0.88 & 13
&  & & & 2$\times10^{-7}$\\ 
& 40.13$\pm$29.05 & Brems. & 1.44$\pm$1.18
& - & - & & 0.88 & 13 &  & & & \\ 
\hline 
9$^m$ & 0.39 & PL & - &
1.61$\pm$0.18 & - & & 0.58 & 20 & 0.7 & 21242 & 1 & 9$\times10^{-3}$/\\ 
& 0.39 & BB & 0.55$\pm$0.06 & - & - & & 1.30 & 20 &  & & &
5$\times10^{-5}$\\ 
& 0.39 & Brems. & 7.99$\pm$4.72 & - & - & & 0.63 &
20 &  & & & \\ 
\hline 17$^m$ & 0.39 & PL & - & 2.37$\pm$0.19 & - & &
1.48 & 19 & 0.80  & 21231 & 1 & 7$\times10^{-2}$/ \\ 
& 0.39 & BB &
0.36$\pm$0.03 & - & - & & 2.79 & 19 &  & & & 8$\times10^{-1}$\\ 
& 0.39
& Brems. & 1.62$\pm$0.37 & - & - & & 1.80 & 19 &  & & & \\ & 0.39 &
BB+PL & 0.10$\pm$0.03 & 1.91$\pm$0.34 & - & & 1.43 & 17 &  & & & \\
\hline 7$^m$ & 0.39 & PL & - & 0.69$\pm$0.55 & - & & 1.22 & 6 & 0.6  &
21231 & 1 & 8$\times10^{-4}$/ \\ 
& 0.39 & BB & 1.31$\pm$0.47 & - & - &
& 1.35 & 6 &  & & & 8$\times10^{-18}$\\ 
\hline 
11$^m$ & 0.39 & PL & -
& 2.02$\pm$0.19 & - & & 0.61 & 19 & 1.1 & 21222 & 1 &
1$\times10^{-2}$/ \\ & 0.39 & BB & 0.37$\pm$0.03 & - & - & & 1.48 & 19
&  & & & 2$\times10^{-6}$\\ & 0.39 & Brems. & 2.46$\pm$0.79 & - & - &
& 0.76 & 19 &  & & & \\ & 0.39 & RS+RS$^@$  & 0.18,4.45 & - & & & 0.52
& 17 &  & & & \\ \hline 22$^m$ & 0.39 & PL & - & 1.66$\pm$0.33 & - & &
0.80 & 9 & 0.8  & 21067 & 1 & 6$\times10^{-2}$/  \\ & 0.39 & BB &
0.40$\pm$0.07 & - & - & & 1.64 & 9 &  & & & 1$\times10^{-1}$\\ \hline
6 & 34.92$\pm$29.00 & PL & - & 0.93$\pm$0.94 & - & & 1.17 & 15 & 2.2 &
21238 & 1 & 6$\times10^{-3}$/ \\ & 4.00$\pm$10.80 & BB & 2.80$\pm$1.32
& - & - & & 1.12 & 15 &  & & & 2$\times10^{-3}$\\ \hline 5 & & & - & -
& - & & -  & -  & 0.3  & 21234 & 1 & 5$\times10^{-3}$/ \\ &  &  &  & -
& - & &  &  &  & & & 5$\times10^{-2}$\\ \hline

\hline

\end{tabular}

PL = power law, BB = blackbody, Brems. = bremsstrahlung, RS = Raymond
Smith\\ $^@$ Errors are $\pm$0.04 and $\pm$1.50 respectively
\end{center}

\end{minipage}
\end{table*}

\begin{table*}
\begin{minipage}{18cm}
\caption{The same as for Table~\ref{tab:xspectram55} except for the
sources in \object{NGC 3201}.  Source 53 uses only the MOS data.  The
N$_H$ of the cluster is \citep[1.17$\times 10^{21}{\rm
cm}^{-2}$][]{harr99}.  Source numbers correspond to those given in
Table~\ref{tab:xsourcesngc3201}.}
\label{tab:xspectrangc3201}
\begin{center}
\begin{tabular}{cccccccccc|ccc}
\hline Src & N$_H$  & Model & kT (keV) & Photon & Abundance & z &
$\chi^{\scriptscriptstyle 2}_{\scriptscriptstyle \nu}$ & dof & Flux &
bins & time & KS/ \\ & $\times 10^{21}{\rm cm}^{-2}$ & & &Index  & & &
& &  & & (s) & $\chi^2$ \\ \hline \hline 26$^m$ & 1.17 & PL & - &
1.90$\pm$0.32 & - & & 1.79 & 14 & 0.3 & 28038 & 1 &
2$\times10^{-5}$/\\ & 1.17 & BB & 0.50$\pm$0.08 & - & - & & 2.18 & 14
&  & & & 2$\times10^{-3}$\\ & 1.17 & Brems. & 3.53$\pm$2.48 & - & - &
& 1.85 & 14 &  & & & \\ & 8.30$\pm$7.50 & RS & 0.97$\pm$1.24 & - &
1.00 & 0.13$\pm$0.25 & 3.11 & 12 &  & & & \\ \hline 23$^m$ &
1.29$\pm$0.34 & PL & - & 1.56$\pm$0.09 & - & & 1.81 & 44 & 2.0 & 28005
& 1 & 3$\times10^{-1}$/ \\ & 0.83$\pm$0.26 & Brems. & 13.28$\pm$4.33 &
- & - & & 1.83 & 44 &  & & & 1$\times10^{-1}$\\ \hline 22$^m$ & 1.17 &
PL & - & 1.54$\pm$0.26 & - & & 0.98 & 9 & 0.4  & 28046 & 1 &
9$\times10^{-3}$/ \\ & 1.17 & BB & 0.60$\pm$0.08 & - & - & & 1.12 & 9
&  & & & 2$\times10^{-1}$\\ & 1.17 & Brems. & 8.15$\pm$7.42 & - & - &
& 0.94 & 9 &  & & & \\ \hline 19$^*$ & 1.17 & PL & - & 2.02$\pm$0.21 &
- & & 1.48 & 17 & 0.3  &  27993 & 1 & 6$\times10^{-1}$/ \\ & 1.17 & BB
& 0.36$\pm$0.04 & - & - & & 2.17 & 17 &  & & & 3$\times10^{-4}$\\ &
0.11$\pm$0.77 & Brems. & 4.94$\pm$3.49 & - & - & & 1.57 & 16 &  & & &
\\ & 0.20$\pm$0.61 & RS & 3.57$\pm$1.14 & - & 1.00 & -  & 1.43 & 16 &
& & & \\ & 1.17 & BB+PL & 0.02$\pm$0.00 & 2.04$\pm$0.24 & - & & 1.67 &
15 &  & & & \\ \hline 16$^m$$^+$ & 1.29$\pm$0.77 & PL & - &
1.87$\pm$0.24 & - & & 1.24 & 25 & 1.0 & 28530 & 1 & 1$\times10^{-3}$/
\\ & 1.17 & BB & 0.40$\pm$0.03 & - & - & & 2.36 & 26 &  & & &
1$\times10^{-2}$\\ & 1.17 & Brems. & 4.17$\pm$1.12 & - & - & & 1.33 &
26 &  & & & \\ & 8.52$\pm$1.87 & RS & 1.03$\pm$0.13 & - & 1.00 & - &
2.90 & 25 &  & & & \\ & 1.17 & BB+PL & 0.25$\pm$0.06 & 1.36$\pm$0.46 &
- & & 1.22 & 24 &  & & & \\ & 1.17 & RS+RS & 6.79+1.04$^@$ & & - & &
1.23 & 24 &  & & & \\ \hline 9$^m$ & 1.17 & PL & - & 1.76$\pm$0.15 & -
& & 0.89 & 27 & 1.3  & 28098 & 1 & 2$\times10^{-5}$/  \\ & 1.17 & BB &
0.44$\pm$0.04 & - & - & & 1.78 & 27 &  & & & 2$\times10^{-2}$\\ & 1.17
& Brems. & 5.33$\pm$2.29 & - & - & & 0.93 & 27 &  & & & \\ &
9.78$\pm$1.91 & RS+RS & 4.82+0.17$^1$ & & - & & 0.73 & 24 &  & & & \\
\hline 33 & 1.17 & PL & - & 1.50$\pm$0.11 & - & & 1.31 & 24 & 1.1 &
28050 & 1 & 9$\times10^{-4}$/ \\ & 1.17 & BB & 0.44$\pm$0.03 & - & - &
& 3.79 & 24 &  & & & 1$\times10^{-4}$\\ & 1.17 & Brems. &
18.26$\pm$14.60 & - & - & & 1.47 & 24 &  & & & \\ & 1.17 & MEKAL &
108.9$\pm$45.5 & - & 2.47$\pm$1.27 & 0.69$\pm$0.07 & 1.37 & 22 &  & &
& \\ \hline 53 & 3.16$\pm$2.28 & PL & - & 2.31$\pm$0.63 & - & & 0.64 &
13 & 2.0 & 33559 & 1 & 1$\times10^{-7}$/ \\ & 1.17 & BB &
0.45$\pm$0.05 & - & - & & 0.97 & 14 &  & & & 2$\times10^{-1}$\\ &
1.79$\pm$1.50 & Brems. & 3.64$\pm$2.89 & - & - & & 0.71 & 13 &  & & &
\\ & 1.67$\pm$1.33 & RS & 4.32$\pm$4.22 & - & 1.0 &  & 0.65 & 13 &  &
& & \\ \hline 13 & & & - & - & - & & -  & -  &  6.0 & 28019 & 1 &
4$\times10^{-5}$/ \\ &  &  &  & - & - & &  &  &  & & &
1$\times10^{-3}$\\ \hline

\hline

\end{tabular}

PL = power law, BB = blackbody, Brems. = bremsstrahlung, RS = Raymond
Smith\\ $^@$ Errors are $\pm$4.54 and $\pm$0.31 respectively.  $^1$
Errors are $\pm$2.92 and $\pm$0.08 respectively \\ $^*$ Neutron Star
Atmosphere (NSA) model, log(T)=6.61$\pm$0.04 K, Mass = 1.4 M$_\odot$
(frozen), Radius = 6.96$\pm$5.51 km, $\chi^{\scriptscriptstyle
2}_{\scriptscriptstyle \nu}$ = 2.09 (16 d.o.f.) \\ $^+$ NSA,
log(T)=6.83$\pm$0.03 K, Mass = 1.4 M$_\odot$ (frozen), Radius =
5.06$\pm$0.26 km, $\chi^{\scriptscriptstyle 2}_{\scriptscriptstyle
\nu}$ = 1.33 (25 d.o.f.) \\ $^+$ NSA+PL, log(T)=6.30$\pm$0.12 K, Mass
= 1.4 M$_\odot$ (frozen), Radius = 10 km (frozen),
$\Gamma$=1.04$\pm$0.63, $\chi^{\scriptscriptstyle
2}_{\scriptscriptstyle \nu}$ = 1.21 (24 d.o.f.)
\end{center}

\end{minipage}
\end{table*}

\begin{figure}
     \includegraphics[angle=0,width=9cm]{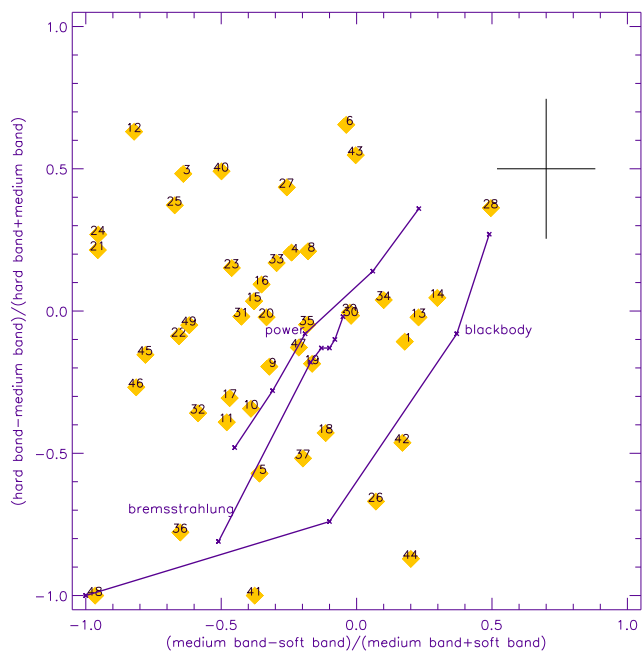} \caption{An
     X-ray colour diagram of the detected X-ray sources in M~55.
     Diamonds with a number indicate an X-ray source where the number
     is that given in Table~\ref{tab:xsourcesm55}. The hard band is
     3.0-10.0 keV, the medium band is 1.5-3.0 keV and the soft band is
     0.5-1.5 keV.  Three lines are indicated showing the colours of a
     hypothetical source in the cluster with a power law spectrum, a bremsstrahlung
     spectrum and a blackbody spectrum.  On the power law spectrum
     line, the crosses (from bottom to top) represent photon indices
     of 2.5, 2.0, 1.5, 1.0 and 0.5.  On the bremsstrahlung spectrum
     line, the crosses (from bottom to top) represent temperatures of
     1.0, 5.0, 10.0, 15.0 and 30.0 keV.  On the blackbody spectrum
     line, the crosses (from bottom to top) represent temperatures of
     0.1, 0.5, 1.0, 1.5 keV.  For N$_H$ values greater than that of
     the cluster, the spectral fits move towards the top right.  A
     typical error bar is also shown in the top right hand corner.}
     \label{fig:xhardm55}
\end{figure}

\begin{figure}
     \includegraphics[angle=0,width=9cm]{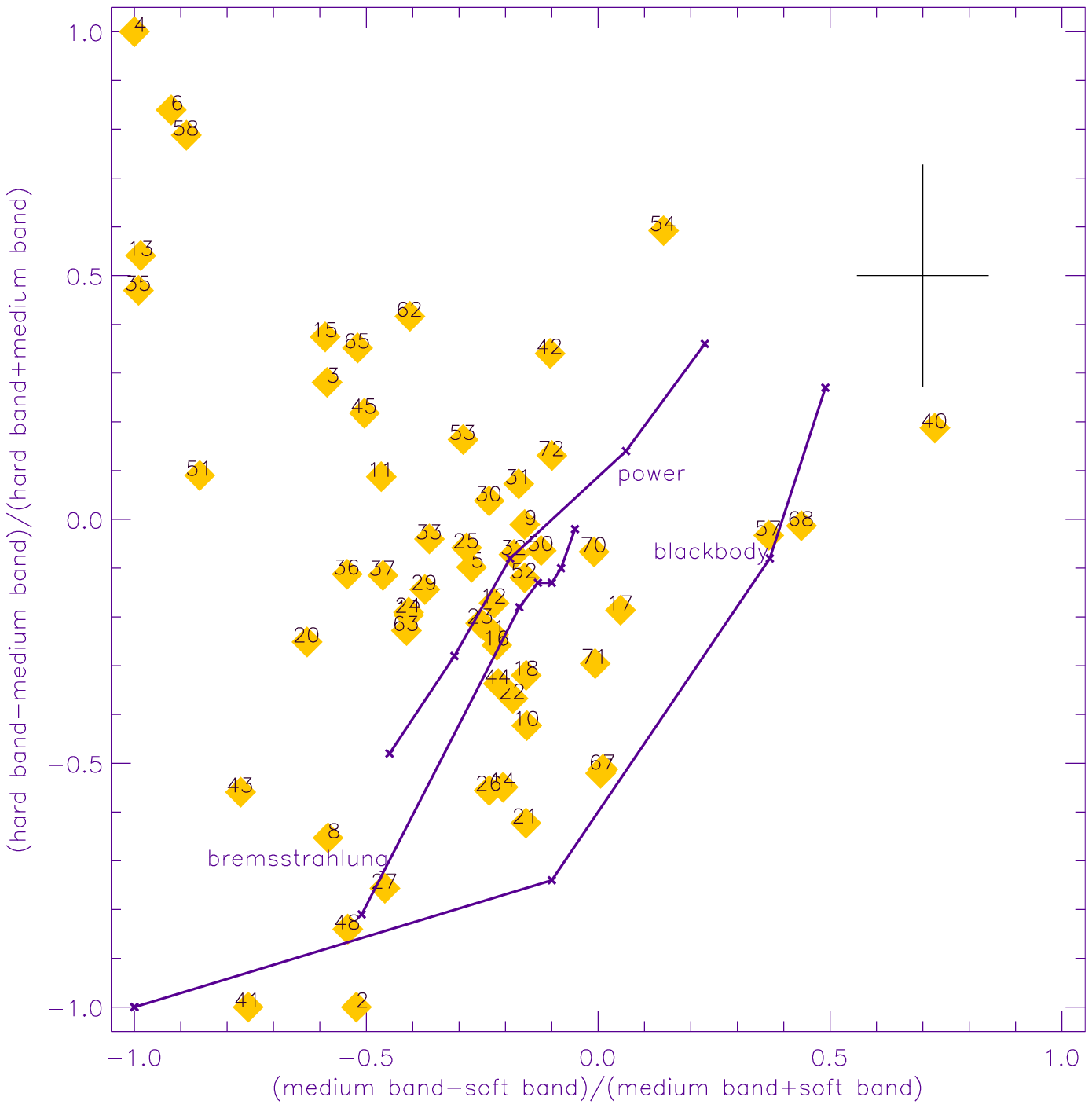}
     \caption{As Fig.~\ref{fig:xhardm55}, but for \object{NGC 3201}.
     The only difference in the symbols is that on the bremsstrahlung
     spectrum line, the crosses (from bottom to top) represent
     temperatures of 1.0, 5.0, 10.0, 20.0 and 50.0 keV.  }
     \label{fig:xhardngc3201}
\end{figure}

As the rest of the sources have insufficient counts to extract a
spectrum, we have used colour-colour diagrams (see
Figs.~\ref{fig:xhardm55} and \ref{fig:xhardngc3201}) to derive
information about their spectral form.

\subsection{Variability analysis}
\label{sec:variability}

We carried out variability tests on all the sources by dividing the
data into two equal frames (10 ks for \object{M 55} and 20 ks for
\object{NGC 3201}) and then into four equal frames (5 ks for \object{M
55} and 10 ks for \object{NGC 3201}).  We looked for variability, 
defined by a change in the frame to frame count rate by more than the
statistical error for each frame (calculated by the source detection
chain).  Using this method we found variable sources in both
clusters.  Sources 22 and 37 in \object{M 55} are variable.  
Using the $\chi^2$ probability of constancy test, these two sources
are variable at the 1 and 3 $\sigma$ level respectively. In
\object{NGC 3201}, sources 42, 45, 55 and 63 are variable.  Again
using the $\chi^2$ probability of constancy test, these sources are
variable at the 4, 2, 2 and 2 $\sigma$ levels respectively.

\begin{figure}
     \includegraphics[angle=270,width=8cm]{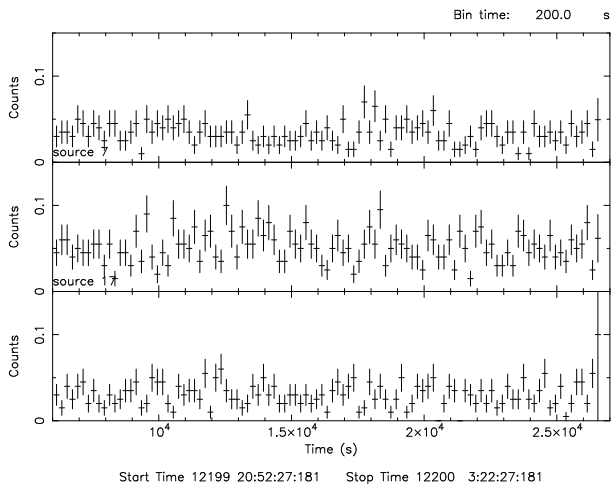}
     \caption{Lightcurves of two of the sources in M55, top: source 7
     and middle: source 17.  The lower panel is a typical background
     lightcurve for these two sources.  All of the lightcurves have
     been plotted on the same scale for ease of comparison and the bin
     size is 200 seconds, as indicated in the top right hand corner.
     The first 6000 seconds of this observation were affected by high
     background and hence are not analysed or plotted.}
  \label{fig:m55lcsa}
\end{figure}

\begin{figure}
     \includegraphics[angle=270,width=8cm]{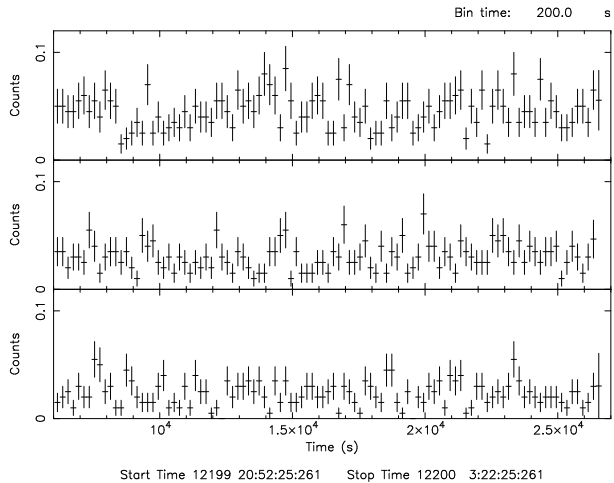}
     \caption{Lightcurves of two further sources in M55, top: source 11
     and middle: source 22.  Again
     the lower panel is a typical background lightcurve for these two
     sources.  All of the lightcurves have been plotted on the same
     scale for ease of comparison and the bin size is again 200
     seconds.  As before, the first 6000 seconds of this observation
     were affected by high background and hence are not analysed or
     plotted.}
  \label{fig:m55lcsb}
\end{figure}

We also carried out a different variability analysis on the sources
given in Tables~\ref{tab:xspectram55} and \ref{tab:xspectrangc3201}.
We extracted the lightcurves using regions of the same size as those
in the spectral analysis (Sect.~\ref{sec:xspect}).  We used the
filtered data between 0.2-10.0 keV and the corresponding GTI (Good
Timing Interval) file.  This is important as gaps in the data can bias
variability tests.  We used the {\it ftool,
lcstats}\footnote{http://heasarc.gsfc.nasa.gov/ftools/} \citep{blac95}
to perform a Kolmogorov-Smirnov probability of constancy test and the
$\chi^2$ probability of constancy test.  The results of these tests,
along with the size and number of data bins are given in
Tables~\ref{tab:xspectram55} and \ref{tab:xspectrangc3201}. Several
sources in each globular cluster  show some evidence for variability.
Of these sources, sources 5, 6, 7 and 22 in \object{M 55} show small
flares in their lightcurve and sources 11 and 17 appear to show some
flickering.  Sources 7 and 17's lightcurves can be seen in
Figure~\ref{fig:m55lcsa} and those of sources 11 and 22 can be seen in
Figure~\ref{fig:m55lcsb}. Using a Kolmogorov-Smirnov test, we
find that the source lightcurve and the corresponding background
lightcurve are statistically different at the 3$\sigma$ level.

We have also looked at the variability between our observations and
previous observations.  \object{M 55} was observed with both the {\em
Rosat} cameras: the PSPC \citep{john96}, and the HRI
\citep{verb01}. \cite{john96} observed the cluster between March and
April 1993 and  found 18 sources in the field of view, where 10 of
these fall in the FOV of our observations.  The flux limit of this
observation is 9.9 $\times 10^{-15} {\rm ergs\ cm}^{-2} {\rm s}^{-1}$
(0.5-2.5 keV).  Only one source was detected within the core/half-mass
radius in both {\em Rosat} observations, our source 30.  This source
has varied in flux by a factor three between the two observations, see
Table~\ref{tab:varM55}.  \cite{verb01} observed the source to have
5$\pm1 \times 10^{-4}$  count s$^{-1}$ (0.5-2.5 keV).  This gives an
unabsorbed flux of 8.1$\pm$1.6 $\times 10^{-14} {\rm ergs\ cm}^{-2}
{\rm s}^{-1}$ (0.5-10.0 keV), consistent with the PSPC observation.
Our source 14 was not detected in either of the {\em Rosat}
observations indicating that it was more than 22 times brighter when
we observed it in 2001 than during the observations made with the {\em
Rosat} PSPC and should have easily been detected by this instrument if
the source had a similar flux in 1993.  We do not detect the
\cite{john96} source 5.  Our flux limit is 2.6$\times 10^{-15} {\rm
ergs\ cm}^{-2} {\rm s}^{-1}$ (0.5-10.0 keV), twenty times fainter than
the source's flux in 1993.  We should therefore have detected this
source in our observations if it had a similar flux in 2001.  This
source must also be variable.  Other variable sources are source 31,
which was a factor 7 times brighter in 1993 and source 11 which was a
factor 9 fainter in 1993, see Table~\ref{tab:varM55}.  Other sources
such as sources 9, 17, 22, 10, 49 and 25 have flux values that do
not vary between the two observations, within the error values, where
the {\em Rosat} errors are of the order 20-25\% of the flux value and
the {\em XMM-Newton} errors bars are 5-25\% of the flux value.

\begin{table}[!h]
\caption{Sources in M~55 found to be variable between these and the
\cite{john96} observations (data taken in March to April 1993).  In
the left hand side of the table, the source number is that given in
Table~\ref{tab:xsourcesm55}.  The unabsorbed flux (${\rm ergs\
cm}^{-2} {\rm s}^{-1}$) is that calculated using the spectrum found
through spectral fitting or through the colour analysis (see
Fig.~\ref{fig:xhardm55}) in the band 0.5-10.0 keV.  In the right hand
side of the table, we give the \cite{john96} count rates in the
0.5-2.5 keV band and their source number (ID).  In the final column,
we calculate the unabsorbed flux (${\rm ergs\ cm}^{-2} {\rm s}^{-1}$)
assuming the same spectral fit as used in the left hand of the table.
This is again quoted in the 0.5-10.0 keV band.}
\label{tab:varM55}
\begin{center}
\begin{tabular}{cc|ccc}
\hline Source  & Unabsorbed & Count rate & ID & Unabsorbed \\ No. &
Flux  ($\times 10^{-14}$) & ($\times 10^{-3}$ cnt s$^{-1}$) & & Flux
($\times 10^{-14}$)\\ \hline \hline 30 & 3.0$\pm0.6$ & 1.55$\pm0.35 $
& 9 & 8.3 $\pm2.0 $  \\

14 & 22.0$\pm$2.0 & 0.0 & - & $<$0.99 \\

-  & $<$0.26 & 1.65$\pm$0.45 & 11 & - \\

31 & 1.0$\pm$0.2 & 3.84$\pm$0.55 & 12 & 7.0$\pm$0.9 \\

11& 18.0$\pm$0.9 & 1.54$\pm$0.40 & 8 & 2.0$\pm$0.5 \\ \hline
\end{tabular}

\end{center}

\end{table}

\object{NGC 3201} was observed by the {\em Rosat} PSPC in December
1991 for approximately 2 ks \citep{john96}.  Only four X-ray sources
were detected in our field of view, our sources 23, 9, 13 and 20.
These sources have not varied between the two observations.

We also searched for periodicities in the barycentre corrected data
using a fourier transform to produce a power density spectrum, with
the {\it ftool, powspec}.  None of the sources showed any strong
evidence for a period.

\section{The optical counterparts}
\label{sec:optcount}

\begin{figure}
\hspace*{-1.2cm}
\includegraphics[angle=0,width=10.5cm]{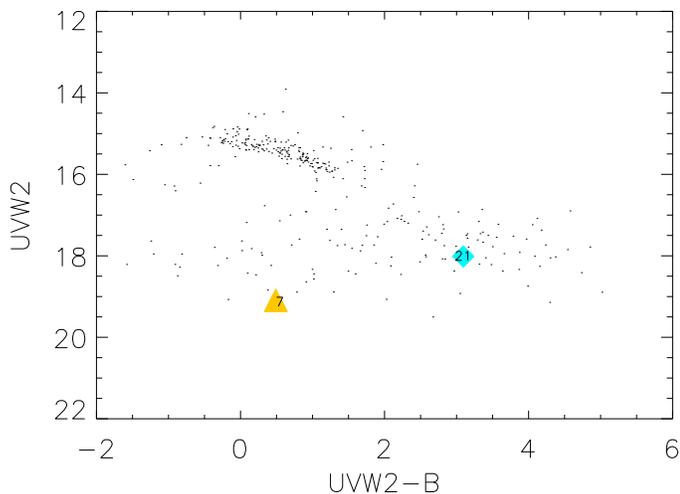}

     \caption{The UVW2, UVW2-B diagram of the globular cluster
     \object{M~55}.  The points show the optical sources and the
     triangle and diamond show the position of the optical couterparts
     to the X-ray source with the identification number given on the
     symbol (as given in Table~\ref{tab:xsourcesm55}).  }
     \label{fig:colormag}
\end{figure}

We constructed a UVW2, (UVW2-B) colour-magnitude diagram with the
optical sources found in \object{M~55}, see Fig.\ref{fig:colormag}.
The majority of the optically blue sources in Fig.\ref{fig:colormag}
are blue giant stars with UVW2 magnitudes between about 15 and 16 and
UVW2-B values between about -0.5 and 1.5.  We have identified two blue
sources that fall in the positional error circles of the X-ray sources and they
can be seen in the colour-magnitude diagram in Fig.\ref{fig:colormag},
where the source numbers correspond to those numbers in
Table~\ref{tab:xsourcesm55}.  Both of these sources lie within twice
the half-mass radius from the centre of the cluster.

Any of the astrometric matches between the {\it XMM-Newton} X-ray
sources and the blue stars may be chance coincidences.  We have
calculated the number of blue stars per unit area as a function of the
radial distance from the centre of the cluster, in the same way as
\cite{edmo03} and \cite{webb04c}. We have thus been able to calculate
the probability of a chance coincidence.  We find probabilities of
10\% and 2\% in the core/half-mass radius and twice the half mass
radius respectively. Therefore, it is unlikely that the two blue sources
that we have detected in the error circle of the X-ray sources are 
chance coincidences, and indeed the optical counterparts to the X-ray
sources.  We do not find an optical counterpart for many other X-ray
sources, but this may be due to the fact that many of the UV sources
in the field of view can not be properly resolved.

To find optical counterparts to the X-ray sources in \object{NGC 3201}
and  to the X-ray sources in \object{M 55} for which we have not been
able to find counterparts in our optical data (as they could be
binaries which have red (cool) optical counterparts), we have also
cross-correlated several catalogues that contain known variable stars
in the two globular clusters \citep{olec99,pych01,mazu03,vonb02}.
None of the variable stars in these catalogues fall within the error
circles of our X-ray sources.  However, \cite{samu96} find a variable
star (their number 51) that is coincident with our source 26 in
\object{NGC 3201}, the most central source of the
cluster. \cite{cote94} undertook a search for spectroscopic binaries
in \object{NGC 3201} using radial velocity variations. They find
sources 215, 216, 248 and 351 respectively \citep{cote94} that fall
within the error circles of our X-ray sources 26, 68 and 63
respectively.  However, they find no evidence for binarity of these
sources.

We also detect several of the optical sources given in the catalogues
above in our optical data of \object{M~55}, where both the positions
and magnitudes are in good agreement with the published data.

\section{Discussion}
\label{sec:discuss}

\subsection{Cluster evolution}

Mass segregation is well known to exist in the majority of clusters,
see \cite{meyl97} for a review.  Certain globular clusters, such as
$\omega$ Centauri, have been found to be exceptions to the rule.  It
is thought that the lack of mass segregation in this cluster could be
due an unusual dynamical evolution of the cluster, through perhaps the
accretion of a stellar system, see \cite{gend03a} and references
therein.

The X-ray sources belonging to the globular clusters are thought to be
mainly degenerate objects and binary systems.  Those sources belonging
to \object{M 55} (see Section~\ref{sec:m55xsources}) are located in
the centre of the cluster, consistent with the idea that the cluster
has undergone mass segregation.  However, the distribution of X-ray
sources in \object{NGC 3201} (see Section~\ref{sec:ngc3201xsources}),
which appear to be evenly dispersed throughout the cluster, up to five
core radii, indicates that the cluster may be slightly
disrupted. \object{NGC 3201} is a slightly unusual cluster as its
orbit is retrograde with respect to the Galaxy \citep{gonz98,vand93}.
It was originally thought that the cluster may have been captured by
our Galaxy, but \cite{gonz98} state that since the eccentricity of
\object{NGC 3201}'s orbit is not yet known, it is possible that it is
an ordinary member of the halo population.  They also remark that
\object{NGC 3201}'s velocity component in the plane of the Milky Way
is none the less unique. \cite{cote95} also find evidence for possible
structure in the velocity field of the stars of the cluster.  They
state that this could be due to several things including the stripping
of stars of the cluster from prolonged interaction with the Galactic
tidal field.  This can perturb the cluster \citep[e.g.][]{gned97},
which could result in the perturbation of the mass
segregation. Indeed, \cite{hein03} state that globular clusters
undergoing disruption should contain an apparent excess of binaries
and binary products, which could explain the apparently large number
of X-ray sources belonging to the cluster.

\subsection{Cluster members}

Here we discuss sources that may be cluster members and their possible
nature.  The following subsections are summarised in
Table~\ref{tab:clustermembers}, where we give all the sources that are
possible cluster members, along with their possible nature and
luminosity if they are cluster members.

\subsubsection{Cataclysmic variables}
\label{sec:discuss_cv}

We find several sources in each cluster which could be cataclysmic
variables (CVs).  CVs have X-ray spectra that are well fitted by high
temperature bremsstrahlung models e.g. \citep{rich96,gend03a}, with
luminosities of up to approximately 1$\times$10$^{32}$ erg s$^{-1}$
(0.5-2.5 keV) \citep{verb97}.  They are variable on short (seconds) to
long (hours/days) timescales. We expect that such objects should be
found towards the centre of the cluster, if the cluster has undergone
mass segregation.

In \object{M 55} source 9 has all of the aforementioned
characteristics.  \cite{verb01} stated that our faint source 30
(his source X9) could be related to the cluster.  It has been noted to
be variable, see Section~\ref{sec:variability}, being 3 times fainter
during our observations than  when it was observed by the {\em Rosat}
PSPC in March 1993 {\em and} by the HRI in September 1997.
\cite{kalu05} observed \object{M 55} over eight observing seasons
spanning the period 1997-2004, in U, B, V and I$_c$.  They observed
six outbursts of a source that they call CV1 and believe to be a dwarf
nova.  CV1 has V= 21.88$\pm$0.06, B-V= 0.63$\pm$0.08,
U-B=-0.83$\pm$0.09 and V-I= 1.18$\pm$0.09 during quiescence and V=
18.98$\pm$0.02, B-V= 0.13$\pm$0.02, U-B=-0.66$\pm$0.03 and V-I=
0.26$\pm$0.03 during outburst.  \cite{kalu05} state that the
variability of CV1 along with its observed colours indicates that it
is a cataclysmic variable of the dwarf nova type.  They also state
that the available data do not allow them to establish with confidence
the cluster membership status of CV1, but that the range of observed
luminosities of the variable, 18 $<$V$<$ 21.8, is consistent with the
assumption that it is a bright, non-magnetic dwarf nova belonging to
M55.  They also indicate that the Rosat source X9 may be the X-ray
counterpart and that the X-ray-to-optical luminosity ratio would be
higher than average for dwarf novae but consistent with that for SS
Cyg, assuming CV1 was in quiescence during {\em Rosat} observations.

CV1 falls within 3.65\arcsec\ of our source 30, well within the
positional error circle of radius 6.98\arcsec\ (see
Table~\ref{tab:xsourcesm55}) and is therefore likely to be the optical
counterpart to source 30.  Looking at the X-ray colour-colour diagram
(see Fig.~\ref{fig:xhardm55}), this source is well fitted by a high
temperature bremsstrahlung model, indicative of a CV. It appears from
our observations that CV1 was not in quiescence during the {\em Rosat}
observations. \cite{kalu05} state that they observed this source in
outburst 23 nights out of the 193 that they observed the cluster, thus
12\% of the time.  If we assume that this is the average time that the
source spends in the outburst state, the chance of observing this
source in outburst twice in two observations is only 1.5\%.  However,
large variations in the X-ray flux of cataclysmic variables have been
observed during quiescence \citep[e.g.][]{bask05}.  If we assume, therefore,
that source 30 was in quiescence during our observations, where the
lightcurve that we extract appears to be of a fairly constant level,
we find that the F${\rm _x}$/F${\rm _{opt}}$= 1.6, where the F${\rm _x}$
value is calculated in the 0.5-2.5 keV band as \cite{edmo03b} and
\cite{verb97}, using a slightly absorbed, 1 keV thermal bremsstrahlung
model and F${\rm _{opt}}$ = 10${\rm ^{-0.4V-5.43}}$. This is still
higher than average for a field dwarf nova, but compatible with a
field magnetic system, see Figure~18 in \cite{edmo03b}.  However,
\cite{edmo03b} discuss the fact that the cataclysmic variables (and
millisecond pulsars) in the globular cluster \object{47~Tuc} have much
higher F${\rm _x}$/F${\rm _{opt}}$ ratios than field CVs and
millisecond pulsars.  They find that the cumulative distribution for
F${\rm _x}$/F${\rm _{opt}}$ of the 47~Tuc CVs follows that of the
non-magnetic, low mass transfer rate field CVs rather than that of the
magnetic CVs and conclude, after considering the optical magnitudes,
that these may be low accretion rate CVs. 

Assuming that the CV is in
the cluster, we find that its luminosity is 5.8$\times$10$^{32}$ ergs
s$^{-1}$, calculated in the 0.1-100 keV band in the same way as
\cite{bask05}, who analysed 34 ASCA observations of non-magnetic
cataclysmic variables. Such a luminosity is greater than that for
their non-magnetic CVs in quiescence, where the most luminous is
SS~Cyg, which has L${\rm _{0.1-100keV}}$=1.2$\times$10$^{32}$ ergs s$^{-1}$.
It is therefore unclear what type of CV source 30 may be, but optical
spectroscopy of this source will help determine its nature.  

In \object{NGC 3201} sources 26 (the most centrally located source,
that has been found to be variable, see Sec.~\ref{sec:optcount}), 23
and 22 all satisfy the above criteria and are likely to be CVs.  The
nature of source 16 is unclear.  Its X-ray spectrum can be equally
well fitted by either a neutron star atmosphere model plus a power law
or a blackbody plus a power law model (see
Table~\ref{tab:xspectrangc3201}) which could indicate that it is a
neutron star system.  However, we do not expect to see such an object
in this low concentration globular cluster (see
Section~\ref{sec:discuss_ns}). Source 16's X-ray spectrum can also be
equally well fitted by two Raymond Smith models (see again
Table~\ref{tab:xspectrangc3201}) indicative of an active binary (see
Section~\ref{sec:discuss_active}).  However, it may also be fitted
using a high temperature bremsstrahlung model, indicative of a CV.

\subsubsection{Neutrons stars}
\label{sec:discuss_ns}

Both quiescent low mass X-ray binaries with a neutron star primaries
(qLMXB$^{NS}$) and millisecond pulsars (MSPs) are expected 
in globular clusters \cite[e.g.][]{gend03b,grin01}.  The qLMXBs$^{NS}$
have luminosities between approximately 10$^{32}$ and 10$^{33}$ erg
s$^{-1}$ and are often well fitted by hydrogen atmosphere models. It
is thought that these objects are formed primarily through collisions,
as opposed to from their primordial binaries, where the number of such
binaries is a function of the collision rate of the cluster
\citep{gend03b,pool03,hein03}.  Due to the low central density and
thus low collision rate of the two globular clusters studied in this
paper, we do not expect to find any such objects.  Millisecond pulsars
can also have soft spectra, that are well fitted by low temperature
blackbodies or hydrogen atmosphere models, where the temperature and
radius of the emission are those of the heated polar cap
\citep[10$^6$-10$^7$ K and $\sim$1 km e.g.][and references
therein]{zhan03,zavl98}.  Millisecond pulsars can also have spectra
that are well fitted by hard power laws, where the emission is due to
particles accelerated in the magnetosphere or spectra composed of the
two different types of emission \cite[e.g.][]{webb04a,webb04b}.  These
objects have lower luminosities than qLMXBs$^{NS}$, unless there is
accretion still taking place in the system.

Sources 7 and 22 in \object{M 55} are well fitted by fairly hard power
law spectra (see Table~\ref{tab:xspectram55}).  However fitting a
blackbody model to these spectra (see again
Table~\ref{tab:xspectram55}) we find approximate temperatures and
radii (source 7, T=1.5$\pm$0.5 $\times$ 10$^7$ K and
r=200$^{\scriptscriptstyle +300}_{\scriptscriptstyle -200}$ m and
source 22, T=4.6$\pm$0.8 $\times$ 10$^6$ K and
r=305$^{\scriptscriptstyle +143}_{\scriptscriptstyle -84}$ m) that are
consistent with those expected from MSPs.  Their interstellar
absorptions are consistent with those of the cluster and their
luminosities are consistent with those expected for millisecond
pulsars and thus they could be MSPs.  We have also found a possible
optical counterpart to source 7, see Fig.~\ref{fig:colormag}.  This is
a very blue source, however it is too bright to be the optical
counterpart of the MSP, unless the MSP has a companion.  Many
millisecond pulsars have companions \citep[e.g.][]{lund96}, but they
can also be cool red stars, rather than the hot blue stars which our
optical dataset is designed to detect.  However, the companion may be
close enough that some accretion onto the compact object continues, in
the same way as in the accreting millisecond pulsar \object{SAX
J1808.4-3658} \citep{mars98}, which could account for the blue nature
of the counterpart.  Alternatively, the companion star maybe
constantly illuminated by the pulsar, as in \object{PSR 1957+20}
\citep{fruc88}, which could also account for the blue nature of the
counterpart.  From their position in Fig.~\ref{fig:xhardm55} and
location in the cluster, sources 42 and 26 could also be MSPs.  Source
17 has a spectrum that is best fitted by a blackbody plus a power law.
Such spectra have been found to fit qLMXBs$^{NS}$ see
\citep[e.g.][]{camp03}. The luminosity (L$_{0.5-10.0
keV}$=3.4$\times$10$^{32}$ erg s$^{-1}$) is also consistent with such
a hypothesis, even though we do not expect any such objects in this
globular cluster.

There are several sources in \object{NGC 3201} that have soft X-ray
spectra (see Fig.~\ref{fig:xhardngc3201}).  However, if this cluster
has not undergone mass segregation, we can not constrain the nature of
the objects through their position, thus sources 21, 27, 48 and 67 may
all be MSPs.  As stated above, the nature of source 16 (in \object{NGC
3201}) is unclear.  It also has a spectrum that is well fitted by
either a blackbody plus a power law or a NSA (Neutron Star Atmosphere)
model plus a power law.  Again, even though we, statistically
speaking, do not expect any such objects in this globular cluster,
this object may also be a qLMXB$^{NS}$.

\subsubsection{Active binaries}
\label{sec:discuss_active}

\cite{demp93} found, following the analysis of 44 active (RS~CVn)
binaries observed with {\it Rosat}, that a two temperature thermal
plasma model, with typical temperatures of 2 $\times$ 10$^6$ K (with a
scatter of approximately 1-3 $\times$ 10$^6$ K) and 1.6 $\times$
10$^7$ K (with a scatter of approximately 1-4 $\times$ 10$^7$ K) gave
a good description of the RS~CVn X-ray spectra. These binaries have
similar luminosities to those of CVs.

We find one source in each cluster which, from the luminosities and
X-ray spectra, appear to be RS~CVn binaries.  Source 11 in \object{M
55} is well fitted by a two temperature Raymond Smith model with
temperatures of 2.1$\pm$0.5$\times$ 10$^6$ K and 5.2$\pm$1.7$\times$
10$^7$ K, with a luminosity consistent with that expected for such
active binaries (see Table~\ref{tab:clustermembers}).  Source 9 in
\object{NGC 3201} is also well fitted by a two temperature Raymond
Smith model with temperatures of 2.0$\pm$0.9$\times$ 10$^6$ K and
5.6$\pm$3.4$\times$ 10$^7$ K and a similar luminosity to the previous
RS~CVn candidate.  Both of the sources show some evidence for
variability in their lightcurves, see Tables~\ref{tab:xspectram55} and
\ref{tab:xspectrangc3201}, which can also be indicative of magnetic
activity on the stellar surface.  Source 16 in this cluster may also
be an active binary from the spectral fitting, see
Table~\ref{tab:xspectrangc3201}, although we see no evidence for
variability in its X-ray lightcurve.

\subsubsection{Other potential cluster members}

At least one of the core/half-mass radius sources in \object{M 55} is
likely to be a member of the cluster, see
Section~\ref{sec:m55xsources}.  Source 30, which appears to be a
cataclysmic variable from the X-ray and optical data is likely to be a
member.  As stated in Section~\ref{sec:discuss_ns}, 42 could be a MSP
and therefore related.  Source 13 may be related to the cluster, from
its position in the cluster and X-ray colours (see
Fig.~\ref{fig:xhardm55}), but it is unlikely that sources 12 and 14
are  members of the cluster (see Section~\ref{sec:nonmembers}).

As stated in Sect.~\ref{sec:discuss_cv}, the most centrally located
source in \object{NGC 3201} (source 26) is likely to be related to the
cluster.  At least one of the core sources are likely to be related to
the cluster, see Section~\ref{sec:ngc3201xsources}.  Source 68, the
has X-ray colours compatible with a source in the cluster (see
Fig.~\ref{fig:xhardngc3201}) and therefore {\it may} also be a CV and
related to the cluster.

\begin{table}[!h]
\caption{Sources which may be related to M55 (top half of the
table) and NGC~3201 (bottom half of the table), along with their
possible nature and their approximate luminosity in the 0.5-10 keV
band ($\times 10^{32} {\rm ergs\ \rm s}^{-1}$) if they belong to the
cluster.  Each list is ordered with the most likely identification to
the least likely identification.}
\label{tab:clustermembers}
\begin{center}
\begin{tabular}{lccc}
\hline Globular & Source & Possible & Luminosity \\ cluster &        &
nature & ($\times 10^{32}$) \\ \hline \hline {\bf M 55} & 30 & CV &
2.2 \\  & 9 & CV & 2.4 \\ & 42 & MSP & 0.4 \\ & 11 & RS CVn & 4.5 \\ &
18 & MSP & 0.8 \\ & 37 & MSP & 0.3 \\ & 17 & qLMXB$^{NS}$ & 3.4 \\ & 7
& MSP & 2.0 \\ & 22 & MSP &  2.7 \\ \hline {\bf NGC 3201} & 26 & CV &
0.9 \\ & 23 & CV & 6.2 \\ & 22 & CV & 1.2 \\ & 9 & RS CVn & 4.1 \\ &
16 & CV/qLMXB$^{NS}$ & 3.1 \\ & 21 & MSP & 0.3 \\ & 67 & MSP & 0.2 \\
& 27 & MSP & 1.1 \\
 
\hline
\end{tabular}
\end{center}
\end{table}

\subsection{Non-cluster members}
\label{sec:nonmembers}

The majority of the sources detected in these observations are
background sources.  In \object{M 55}, source 14 appears to be a
background source, from the high interstellar absorption alone, which
is much greater than that of the cluster.  This is not a defining
criteria for a background source, as certain objects that exist in
globular clusters can have absorption values higher than that of the
cluster i.e. cataclysmic variables.  CV spectra can be absorbed due to
the high intrinsic absorption of the X-rays from the inner disc and/or
white dwarf passing through the edge-on accretion disc e.g. one of
the CVs in \object{M 80} \citep{hein03}. The variability of this
source and the good fit to the spectrum with a low temperature
blackbody could indicate that the source is a low mass X-ray binary
that contains a neutron star.  However it's luminosity in quiescence
is too low for it to be such an object.  It therefore seems likely
that this object is a background object.


Source 6 in \object{M 55} is also likely to be a background object
from its interstellar absorption, spectral fit and position in the
cluster.  Source 28 has an extremely hard spectrum, in fact the
hardest of the cluster.  It is likely that this source is not actually
related to the cluster, but is an Active Galactic Nucleus (AGN). The
fact that we find no optical counterpart to this source in our optical
data could support this theory, where the optical flux would be
redshifted to very red wavelengths.  Other sources that are found in
the top right hand corner of Fig.~\ref{fig:xhardm55}, that are found
towards the exterior of cluster, such sources 27 and 40 are also
likely to be background objects.

\cite{kalu05} have detected a blue source in their observations
(described in Section~\ref{sec:discuss_cv}), which they call
\object{M55-B1}, which has V= 20.40 and U-B=-0.65. This source falls
within 2\arcsec\ of our source 39, well within the error circle of
radius 8\arcsec.  \object{M55-B1} is therefore likely to be the
optical counterpart to source 39.  From its colours and low-amplitude
variability, \cite{kalu05} propose that this source is a quasar.
Source 39 does not appear in the X-ray colour-colour diagram of
\object{M 55} (Figure~\ref{fig:xhardm55}).  Extracting the spectra, we
find that the source is very absorbed and has insufficient counts in
the softest band to appear on the diagram.  We find that the
interstellar absorption is approximately 1$\times$10$^{23}$ cm$^{-2}$,
which could indicate an extragalactic origin for this object.

In \object{NGC 3201}, the spectra of sources 19 and 33 are best fitted
by a Raymond Smith and a MEKAL model respectively, both indicative of
emission from a hot diffuse system, such as a galaxy.  They therefore
could be extragalactic sources.  Sources 4 and 6 are located far from
the cluster centre and have spectra consistent with that of an AGN. Source
58 in the half-mass radius, could belong to the cluster (see
Section~\ref{sec:ngc3201xsources}), however from the X-ray colours,
which indicate an absorbed extragalactic source, it is unlikely.
Other sources with similar colours such as 6, 13 and 54 may also be
background sources.

\section{Conclusions}

Using {\it XMM-Newton} X-ray observations of the globular \object{M
55} we detect 47 X-ray sources in the direction of the cluster, any of
which belonging to the cluster must have low luminosities.  Using the
logN-logS diagram derived from Lockman Hole observations to determine
the background population, we find that very few of the sources
(1.5$\pm$1.0) in the FOV of this cluster are likely to belong to the
cluster.  These sources are found in the central region of the
cluster, which implies that this cluster has undergone at least some
mass segregation.  Through X-ray and optical colours, spectral and
timing analysis (where possible) and comparing these observations to
previous X-ray observations, we find two sources that could be
cataclysmic variables, one that could be an active binary, several
that may be MSPs and possible evidence for a qLMXB$^{NS}$.  The
majority of the other sources are background sources such as AGN.

We find 62 X-ray sources in the {\it XMM-Newton} FOV of \object{NGC
3201}, where as many as 15 sources could belong to the cluster.  These
sources, in contrast to those found in \object{M 55} and in many other
Galactic globular clusters are not centrally located, but distributed
throughout much of the cluster, which could indicate a perturbation of
the cluster.  Using X-ray observations only, we find three/four
sources that could be CVs, one that could be an active binary, several
that may be MSPs and again possible evidence for a qLMXB$^{NS}$,
although again we do not expect any such objects, statistically
speaking, in either of the clusters, due to their low central
concentrations.  Again, the majority of the other sources are
background sources.

\begin{acknowledgements}

This article was based on observations obtained with XMM-Newton, an
ESA science mission with instruments and contributions directly funded
by ESA Member States and NASA.  The authors NAW and DB also
acknowledge the CNES for its support in this research.

\end{acknowledgements}

\end{document}